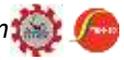

# A Review of Augmented Reality Apps for an AR-Based STEM Education Framework

**Carlo H. Godoy, Jr.**
Philippine Navy
godoy.carlo@navy.mil.ph

**Abstract**

Within the past two decades, Augmented Reality (AR) applications have received increased attention. Augmented Reality is now widely used in the education sector at level K-12. AR is expected to be generally adopted in two-to-three years in higher education and four-to-five years in K-12. Applying AR technology in the education sector especially in STEM subjects, can result in having a smart campus. In adopting a SMART Campus strategy, education practitioners must address many intrinsic issues in science, technology, engineering, and mathematics (STEM) research. For example, in physics, there are expensive or insufficient laboratory systems, system faults, and difficulty simulating other experimental circumstances; in technology, many schools do not have enough computers; in engineering, there are only a few instructors who are knowledgeable in computer-aided design (CAD); and in mathematics, few teachers incorporate technology into their lessons often because they believe it is still better to teach through the traditional methods. Hence, In this paper we discuss how AR is being used now in different learning areas in STEM to open new doors to researchers and teachers as they transition their schools into SMART campuses with the use of AR apps. Aligned with this, a suggested framework for school administrators and policymakers is proposed based on a review of the positive benefits of different AR apps.

*Author Note: A Review of Augmented Reality Apps for an AR-Based STEM Education Framework is an extension study stemming from a previous review that the author conducted. The extension study was done to craft a framework for STEM Education that the author hopes will help schools in the Philippines.*

Augmented Reality (AR) apps have received increased attention over the previous two decades. AR generates fresh world experiences with its data layering over 3D space, suggesting that AR should be embraced over the next 2–3 years to give fresh possibilities for teaching, learning, study, or creative investigation according to the 2011 Horizon Report (Chen et al., 2017). This article discusses the different augmented reality applications that are being used in STEM education and will then create a suggested framework based on the review results. Azuma, as cited by Akçayir and Akçayir (2016), states that virtual objects in an Augmented Reality applications appear in coexistence in the same space as the objects that are located in the real world, hence it is very well-suited for enhancing STEM learning environments. AR is now a common technology used in instructional environments in the education sector (Fernandez, 2017).

According to Rouse, as cited in Jung and tom Dieck (2018), AR is the integration of information in digital format, which includes live video in the real time environment of a certain user. In augmentation of live videos, integrating a video picture to a digital environment involves identification of an object replicated from features of the physical world. It is then captured as any format that will be considered as a video picture. It will mean increasing the responsiveness of the generated video picture to the state needed to control the object from the physical world itself (Kochi et al., 2017). In an augmented reality system, the integrated digital information can only be seen using devices like phone cameras. Hence, it cannot be seen in the real world without using a camera. These digital information forms can be represented in different forms like a stack of virtual cubes or manipulating a non-real object in a variety of ways (Hilliges et al., 2018). AR can also be used to indicate or layer additional information onto a real environment. This supplemental information is considered optional and may not affect the actual user of the system itself. The methods being used by an AR system to provide these supplemental information are the following: (a) Tracking the user's point of view, (b) Capturing a camera field of perspective, and (c) Obtaining additional data in the field of perspective captured for at least one object (Pasquero & Bos, 2017). One example of supplemental information is found in vehicles. In this manner, it obtains additional data in the field of a perspective captured for at least one object. Hence, if the user's interest has been captured, the system should present an augmented reality replica of a vehicle and cover the user's environment based on the point of interest (Habashima et al., 2017).

**Augmented Reality for STEM Education**

AR is widely used now in the K-12 level of the education sector (Akçayır & Akçayır 2016). Ferrer-Torregrosa et al. (2015) stated that AR is also being used now by different universities. The application of the technology in the education sector can lead to a "smart campus." Smart campuses are designed to benefit professors and students, handle the resources available and improve the experience of the users with proactive services (Ozcan et al., 2017).

As highlighted in the Horizon report, AR is acknowledged as one of the most significant innovations in Higher Education and K-12 education technology (Johnson et al., 2015). AR is gradually becoming integrated as an emerging technology in the region of inclusive education that adapts learning to provide equal footing through accessible exploration and experiences by all (Marín-Díaz, 2017). Johnson et al. stated that AR is anticipated to be widely adopted in higher education in two-to-three years and in K-12 in four-to-five years (cited by Saltan, 2017). Consequently, it is essential to explore how teachers and scientists incorporate AR into teaching-learning procedures if this is the present state of the art for the use of AR in education. AR became visible in the early 2000s and its effectiveness for learning was soon established by educational research (Dede et al., 2017).

**Review Methodology**

To share more about the ways in which AR is being used in STEM education, the author conducted a focused review of AR in STEM education research. A representation of the strategy employed in completing the review is shown in Figure 1. The search procedure started by selecting the topic to be reviewed. In this case the topic is AR for STEM Education. The topic selected explored the different sectors in STEM that are using Augmented Reality as a tool for teaching and learning. After identifying the topic, the next step was to search Google Scholar. About 1,530,000 results were noted with 97,600 filtered to explore further. In this part the study was filtered depending on the importance of each study. The importance was filtered using exclusion-inclusion criteria, which are shown in Table 1. Another filter that needed to be added is the year when the publication was published. It assumed that a five year interval will still make a certain publication valid. Once the filtering was been set, the author selected the relevant documents that would build the foundation of the review. This study will only focus on how the different disciplines are being used in the education sector and is not deemed as a comprehensive review. After knowing the foundation and the outline where the documents would be discussed, the DOI (Digital Object Identifier) of each publication was used to be able to access the full copy. Once the full study had been acquired, all documents were reviewed. During the review process, it was necessary to filter which documents were needed to support the selected topic. Once the important facts had been gathered and the studies had been filtered, the review was written.

*Figure 1*

*Review Methodology*

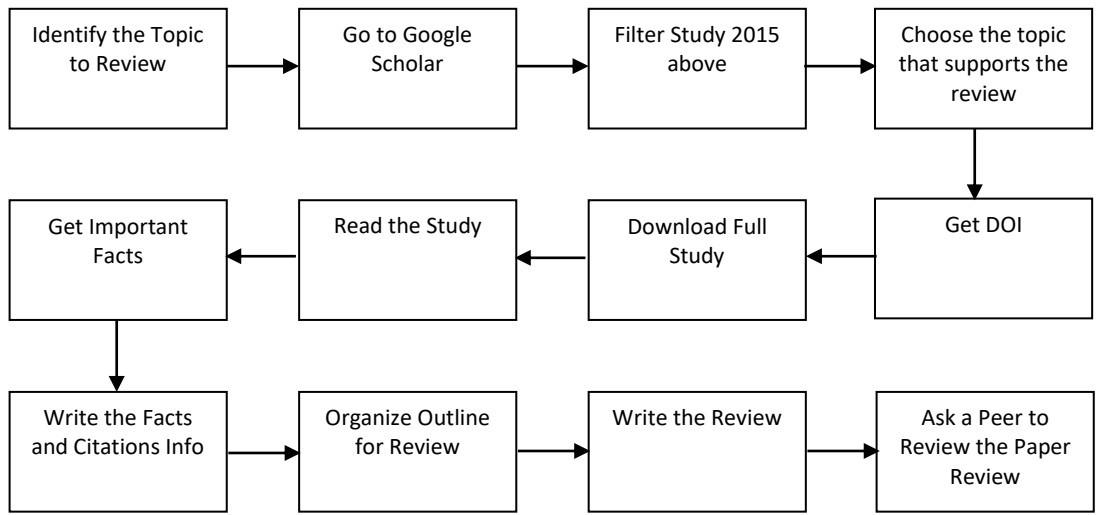

**Reviewing Augmented Reality for STEM (Science, Technology, Engineering, Mathematics) Education**

After applying the exclusion and inclusion criteria (Table 1), the author selected different literatures that helped in the discussion of the review.

*Table 1*

*Exclusion and Inclusion Criteria*

| Exclusion Criteria | Inclusion Criteria |
|---|---|
| Comparison of GBL to normal and blended learning | Effectiveness of games for educational purpose as a supplementary tool |
| Games that are not for SPED, Language Skills and Technical Vocational Education | The study should either be for SPED, Language Skills or Technical Vocational Education |
| Duplicate Studies | The game should be visually impacting in terms of the subject area |

There was no definite number of articles included but the author has chosen more than 10 literature reviews and included the most suitable for the review. Below are the results per discipline.

**Science**

Education Professionals must tackle several problems intrinsic in the training of science fields such as physics costs or inadequate laboratory equipment, mistakes of equipment, or difficulty in simulating certain experimental circumstances (Cai et al., 2017). These normally lead to student's learning achievement in physics to be lower, hence students may have low motivation in learning this subject. Augmented Reality can be a successful approach to tackling these problems. A study about magnetic field instruction has been conducted in relation to the problems. Results of the analysis demonstrated that the movement-sensing software based on AR can enhance the learning attitude and learning result of the learners. This research offers a case for applying AR technology to secondary education in physics (Cai et al., 2017). In learning about health science, medical anatomy, and neurosurgical procedures it is also very helpful to use Augmented Reality as a learning tool. In an environment where a required structure needs to be examined from all angles, anatomical learning is best performed using a tool that will show these angles (Moro et al., 2017). Augmented Reality is one of the best tool to show angles as the developer can easily manipulate how the augmented object will rotate and be displayed.

Compared to traditional pedagogical schemes, Virtual Reality (VR) and AR have the ability to produce improved teaching environments. 3D learning environments can increase the motivation/engagement of learners, improve the representation of spatial information, improve learning contextualization and create superior technical skills (Pelargos et al., 2017). Over the previous several centuries, neurosurgical has experienced a technological revolution, from trephination to image-guided navigation. Advances in Virtual Reality and Augmented Reality are some of the latest ways of integrating into neurosurgical exercise and resident education (Pelargos et al., 2017). Another application named Guided Exploration Training MAR software has been tested. Majid et al. (2021) conducted a research to test this AR application using a post and pretest method and the result has shown a positive feedback. In this study, visualizing each atom's 3D model helps the students to gain more insight into the chemical properties of group one metals. The increase in scores demonstrates that in the end complicated materials can be used and learned using the Guided Exploration Training MAR software.

Studies have shown that AR technology can significantly improve learning results in education. For example, AR enables learners to participate in real-world explorations such as marine life explorations that not everyone has been able to experience (Akçayir et al., 2016). Marine schooling includes problems that are rich and multifaceted. Raising awareness of marine settings and problems requires fresh teaching materials to be developed. In line with that, a digital game-based learning was tailored for primary school learners to design an innovative marine learning program incorporating augmented reality (AR) technology (Lu & Liu, 2015). The results of using this technology are the following: (a) Learners were extremely confident and satisfactorily viewed the learning operations ; (b) Learners obtained target goals for understanding ; and (c) The innovative teaching program specifically helps small academic achievements and enhance learning efficiency. Another great application of Augmented Reality in science is an AR-based simulation scheme for a cooperative investigation-based teaching activity in a science course and which discovered that AR-based simulation could involve learners more deeply in the investigatory project activity than traditional simulations could (Hwang et al., 2016).

**Technology**

Augmented reality has attracted great government attention among these techniques because it offers a fresh teaching view by enabling learners to visualize complicated spatial relationships and abstract ideas (Phon et al., 2015). Research has shown that, owing to a number of factors, many Malaysian non-technical learners have low motivation in studying ICT courses, such as absence of effective teaching practices and efficient teaching apps. In the outlook of such issue, the research teams conducted a quasi-experimental analysis to examine the adverse effect of a new application for mobile augmented reality learning (MARLA) on the motivation of learners to learn a topic of an ICT course at a university (Hanafi et al., 2017). Another study under the ICT education sector has a goal of exploring whether the integration of AR methods would facilitate application for changing the style as well as analyzing a distinct outcome in educating the learners, which uses a blended learning approach based on online methods and AR (Wang, 2017). It was found that technology instructional scientists should take cautious consideration of the educational goal architecture, the data size shown on the cellphone monitor, and the teaching machinery and school facilities setting when incorporating AR apps into a course in order to obtain an appropriate learning situation.

The effect of Augmented Reality and QR Code Integration on achievements and views of undergraduate students taking computer training was examined by Bal and Bicen (2016). A test study group included 50 volunteer students taking a compulsory computer course studying in Near East University's Department of Guidance and Psychological Therapy. The study used experimental research design. Students were divided into two classes at the beginning of the term, namely experimental and control groups, consisting of 25 students in each group. During the first lecture, a pre-test was given to the two groups and a lecture was given using conventional methods, in which computer features were described and computer feature functions were explained to the $1^{st}$ group. Three-dimensional images of hardware features in the computer course hardware chapter were demonstrated to the $2^{nd}$ group by means of augmented reality and QR technology a  laboratory environment with screen, projection, and voice systems. During the lecture, students examined computer hardware with QR code cards and increased reality technology from their own mobile devices, and all lectures were given until the end of the semester in this way.

At the end of the semester, a post-test was given to the experimental and control groups and the experimental group also was provided a questionnaire related to the virtual reality and computer hardware course application with QR code integration. Results from the study showed that the experimental group's level of achievement was higher. Results showed in this context that computer hardware course implementation with augmented reality and QR code integration has a positive impact on students and their academic accomplishments with positive views towards this course. Based on the findings, the approach used in this study is expected to lead to the introduction of technology into education and thus technology and education should be mutually beneficial.

**Engineering**

Augmented reality is very effective in engineering and can have many applications, which includes boosting the learner's motivation in this field. AR may help an engineer or designer to design a product in the right environment, enabling them to be mindful of space limitations or other barriers. It may also encourage an engineer to incorporate esthetics into their design, making sure that when it is done and assembled their product would look pleasing to the eye. Or maybe an engineer may need to build an improvement to an existing component or system, but they do not have the original element drawings or models. Considering the size, shape and existing features, AR allows an engineer to design directly on the existing item (Heimgartner, 2016).

A study conducted by Martin-Gutierrez et al., (2014) has provided a clear proof of the positive used of AR in Engineering Education. Analysis provides findings that demonstrate how engineering students produce better academic outcomes and are more inspired by integrating the latest generation of technical resources into the learning process. Twenty-five first year students studying for a degree in Mechanical Engineering used AR technology to assist them in the graphic engineering subject matter. A control group of twenty-two classmates used conventional class notes during the study. Both these students took an analysis and two surveys to provide input on the teaching content: one to find out the efficacy and quality of the material itself, along with the student satisfaction level; Another for measuring student motivation by using the available technologies during the course of the research. The findings revealed a substantial statistical disparity in their academic outcomes and appeared to be higher in the experimental community; this community also displayed a higher motivation level than the control group.

**Mathematics**

An integrated STEM (Science, Technology, Engineering and Mathematics) lesson requires to participate and nurture students' interest in real-world circumstances, which has been proven to boost learner's motivation in this subject and while real-world STEM situations are naturally incorporated, the embedded STEM contents are rarely taught by school educator (Hsu et al., 2017). One of the hardest subjects of that track is Mathematics. One example of a Mathematics subject is Solid Geometry. To give a better experience in learning solid geometry, a study has been conducted to combine Augmented Reality (AR) technology into teaching operations designing a learning scheme that helps junior high school learners learn sound geometry (Liu et al., 2019; TeKolste & Liu, 2018). Based on the result of the study, AR really gives a big leap in learning solid geometry.

Castillo (2015) on the other hand has introduced a new software framework for the creation of AR applications based on publicly available components. It offers a comprehensive view of the subsystems and the tasks involved in developing a mobile AR application. The standard task of plotting a quadratic equation was chosen as a case study to gain feasibility insights into how AR could help the teaching-learning process, and to analyze the student's reaction to the technology and the application. The pilot study was carried out in a Mexican Undergraduate School with 59 students. To collect information on the experience of the students using the AR application, a questionnaire was developed, and the review of the results obtained is presented. The findings obtained by applying the questionnaire on the use of AR technology indicate that using AR can help enhance the teaching-learning process in Mexican classrooms and inspire the students and can be an innovative tool to revolutionize the learning paradigm.

Another study deals with the use of AR in teaching and learning math that uses this technology to its complete benefit in providing concrete experience in interacting with revolutionary solids. At the end of the study, it was found out that Augmented Reality is beneficial in the understanding of computing solids of revolution volumes (Salinas & González-Mendívil, 2017). AR techniques are strongly linked to calculation capacity and computational calculations, and therefore their evolution is related to personal computer development. It is therefore essential to begin by referring to some of the works that have been created through the implementation of these techniques at global and national level, primarily in the field of education and teaching (Coimbra et al., 2015). It can be inferred that from the birth of AR it was already related to Mathematics. AR makes mathematical ideas simpler to comprehend because it provides better visualization and interaction. We can therefore conclude that three-dimensional techniques, such as AR, improve mathematics teaching and learning. At the same time as the imperative to better comprehend the use of mobile devices for learning mathematics in many nations, there is a powerful political will to enhance teaching and learning process in mathematics education to support innovation that drives economic growth and create the capacity of tomorrow's workers for future work markets (Bano et al., 2018).

**AR-Based Framework For Stem Education Derived From The Review As A Basis For School Administration And Policy Makers**

*Figure 2*

AR-Based STEM Framework

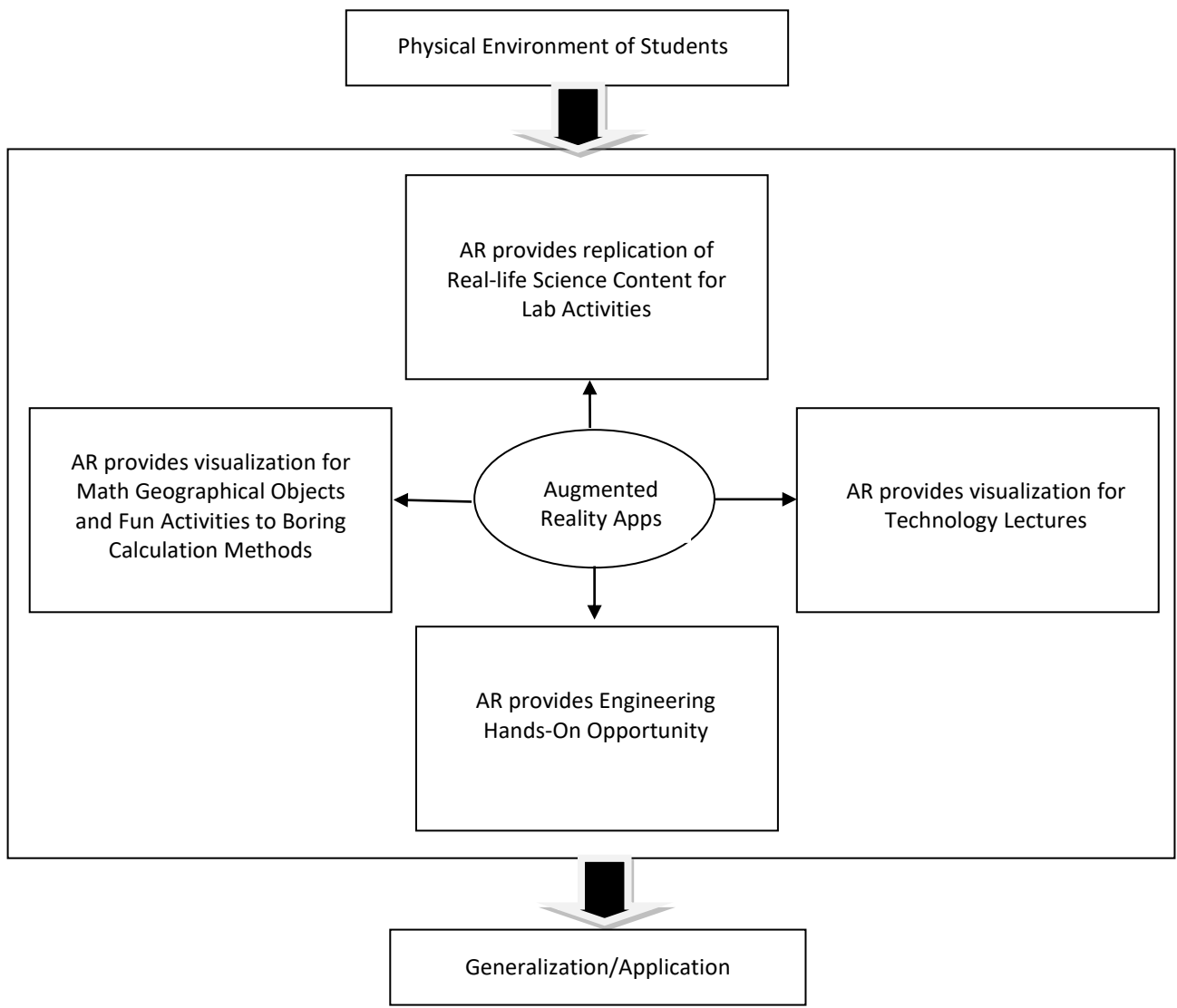

A STEM (Science, Technology, Engineering and Mathematics) education framework can now be derived from the review. STEM lesson requires to participate and nurture students' interest in real-world problems and lectures, so the framework shows that Augmented Reality integrates the physical environment with the content and activities. The different applications discussed on the review has

C. H. Godoy 69

shown that Augmented Reality has been a successful aid in learning the different subject areas under the STEM strand. According to Kelley et al. (2016), enhanced integration of STEM subjects might not be more successful if the implementation strategy is not strategic. A well-integrated curriculum, however, gives students the ability to learn through more meaningful and engaging ways, promotes the use of higher critical thinking skills, enhances problem-solving abilities and increases retention. In that case, the above framework serves a guide for school administrators, policy makers and STEM Department Heads on how AR-Based Activities can be implemented as a supplementary learning tool as part of their blended learning approach. Given this framework, a physics teacher might very well be able to increase the student's learning ability.

## CONCLUSION

As shown in the results of the review, AR has been proven to help organizations and teachers in teaching different STEM disciplines and can be applied here in the Philippines as well as in other areas of the region. Studies have shown that the use of AR can be more efficient in teaching support than other technology-enhanced settings. If content is represented as 3D to learners, it is possible to manipulate objects and handle the information interactively. Rapid technological evolution changed the face of education, especially when technology was coupled with adequate pedagogical foundations. This combination has developed fresh possibilities to enhance the quality of teaching and learning experiences (Nincarean et al., 2013).

Based on the findings, Augmented Reality and the framework presented can help boost the education sector of the Philippines. The combination of these two processes will definitely lead to a new system which will have a major impact on the STEM sector of Education (Phon et al., 2015). In implementing this new system, the AR-Based Stem framework is helpful for school administrators, policy makers and STEM Department Heads

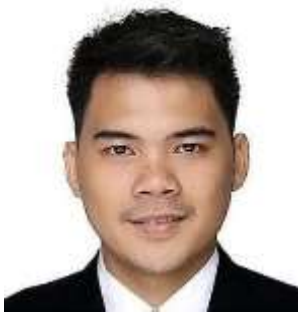

Carlo H. Godoy, Jr. is a former Escalations Manager at Novartis Pharmaceutical and Support Analyst for SQL at Human Edge Software Philippines. Currently he is a Computer Programmer/Sail Plan Manager at the office N6 of the Philippine Navy. He is also a Research Scholar at the Technological University of the Philippines in the Masters in Information Technology program specializing in studies related to educational technology using emerging technologies like augmented reality. Mr. Godoy has been in the education sector for eight years. He currently serves as a thesis adviser in the Technological University of the Philippines, Manuel L. Quezon University, and in STI College Munoz Edsa.

## REFERENCES

Akçayir, M., Akçayir, G., Pektaş, H. M., & Ocak, M. A. (2016). Augmented reality in science laboratories: The effects of augmented reality on university students' laboratory skills and attitudes toward science laboratories. *Computers in Human Behavior, 57*, 334–342. https://doi.org/10.1016/j.chb.2015.12.054

Akçayır, M., & Akçayır, G. (2016). Advantages and challenges associated with augmented reality for education: A systematic review of the literature. *Educational Research Review*. https://doi.org/10.1016/j.edurev.2016.11.002

Bal, E. & Bicen, H. (2016). Computer hardware course application through augmented reality and QR code integration: Achievement levels and views of students. *Procedia Computer Science, 102,* 267-272.

Bano, M., Zowghi, D., Kearney, M., Schuck, S., & Aubusson, P. (2018). Mobile learning for science and mathematics school education: A systematic review of empirical evidence. *Computers and Education*, *121*(February 2017), 30–58. https://doi.org/10.1016/j.compedu.2018.02.006

Cai, S., Chiang, F. K., Sun, Y., Lin, C., & Lee, J. J. (2017). Applications of augmented reality-based natural interactive learning in magnetic field instruction. *Interactive Learning Environments*, *25*(6), 778–791. https://doi.org/10.1080/10494820.2016.1181094

Chen, P., Liu, X., Cheng, W., & Huang, R. (2017). *A review of using Augmented Reality in Education from 2011 to 2016*. 13–18. https://doi.org/10.1007/978-981-10-2419-1

Coimbra, M. T., Cardoso, T., & Mateus, A. (2015). Augmented reality: An enhancer for higher education students in maths learning? *Procedia Computer Science*, *67* (January 2016), 332–339. https://doi.org/10.1016/j.procs.2015.09.277

Dede, C. J., Jacobson, J., & Richards, J. (2017). *Introduction: Virtual, augmented, and mixed realities in education*. https://doi.org/10.1007/978-981-10-5490-7_1

Fernandez, M. (2017). Augmented-Virtual reality: How to improve education systems. *Higher Learning Research Communications*, *7*(1), 1. https://doi.org/10.18870/hlrc.v7i1.373

Ferrer-Torregrosa, J., Torralba, J., Jimenez, M. A., García, S., & Barcia, J. M. (2015). ARBOOK: Development and assessment of a tool based on augmented reality for anatomy. *Journal of Science Education and Technology*, *24*(1), 119–124. https://doi.org/10.1007/s10956-014-9526-4

Habashima, Y., Kurosawa, F., Alaniz, A., & Gleeson-May, M. (2017). *System and method for providing an augmented reality vehicle interface*. US patent:
US9550419B2 - System and method for providing an augmented reality vehicle interface - Google Patents

Hanafi, H. F., Said, C. S., Wahab, M. H., & Samsuddin, K. (2017). Improving students' motivation in learning ICT course with the use of a mobile augmented reality learning environment. *IOP Conference Series: Materials Science and Engineering*, *226*(1). https://doi.org/10.1088/1757-899X/226/1/012114

Heimgartner, J. (2016, April 13). *What Is augmented reality and how can engineers and designers use it? Engineering.com*. Https://Www.Engineering.Com/. https://www.engineering.com/DesignSoftware/DesignSoftwareArticles/ArticleID/11873/What-Is-Augmented-Reality-and-How-Can-Engineers-and-Designers-Use-It.aspx

Hilliges, O., Kim, D., Izadi, S., & Weiss, M. H. (2018). *Grasping virtual objects in augmented reality*. *1*(12). US patent:  US9552673B2 - Grasping virtual objects in augmented reality - Google Patents

Hsu, Y. S., Lin, Y. H., & Yang, B. (2017). Impact of augmented reality lessons on students' STEM interest. *Research and Practice in Technology Enhanced Learning*. US patent: https://doi.org/10.1186/s41039-016-0039-z

Hwang, G. J., Wu, P. H., Chen, C. C., & Tu, N. T. (2016). Effects of an augmented reality-based educational game on students' learning achievements and attitudes in real-world observations. *Interactive Learning Environments*, *24*(8), 1895–1906. https://doi.org/10.1080/10494820.2015.1057747

Johnson, L., Adams Becker, S.., Estrada, V., & Freeman, A. (2015). New media consortium Horizon r eport: 2015 museum edition. In *New Media Consortium*. https://eric.ed.gov/?id=ED559371

Jung, Timothy & tom Dieck, C. M. (2018). Augmented reality and virtual reality: Empowering human, place and business. In *Springer International: Vol. XI* (Issue 5). http://search.ebscohost.com/login.aspx?direct=true&db=a9h&AN=9084262&login.asp&site=ehost-live

Kochi, M., Harding, R., Campbell, D.A., Ranyard, D. & Hocking, I. M. (2017). Apparatus and method for augmented reality. US patent https://patents.google.com/patent/EP2662838B1/en.

Liu, E., Li, Y., Cai, S., & Li, X. (2019). *The effect of augmented reality in solid geometry class on students' learning performance and attitudes: Proceedings of the 15th International Conference on Remote Engineering and Virtual Instrumentation* (Vol. 47, Issue October 2018). Springer International Publishing. https://doi.org/10.1007/978-3-319-95678-7

Lu, S. J., & Liu, Y. C. (2015). Integrating augmented reality technology to enhance children's learning in marine education. *Environmental Education Research*, *21*(4), 525–541. https://doi.org/10.1080/13504622.2014.911247

Marín-Díaz, V. (2017). The relationships between augmented reality and inclusive education in higher education. *Bordón. Revista de Pedagogía*, *69*(3), 125. https://doi.org/10.13042/bordon.2017.51123

Moro, C., Štromberga, Z., Raikos, A., & Stirling, A. (2017). The effectiveness of virtual and augmented reality in health sciences and medical anatomy. *Anatomical Sciences Education*, *10*(6), 549–559. https://doi.org/10.1002/ase.1696

Nincarean, D., Alia, M. B., Halim, N. D. A., & Rahman, M. H. A. (2013). Mobile augmented reality: The potential for education. *Procedia - Social and Behavioral Sciences*, *103*, 657–664. https://doi.org/10.1016/j.sbspro.2013.10.385

Ozcan, U., Arslan, A., Ilkyaz, M., & Karaarslan, E. (2017). An augmented reality application for smart campus urbanization: MSKU campus prototype. *ICSG 2017 - 5th International Istanbul Smart Grids and Cities Congress and Fair*, 100–104. https://doi.org/10.1109/SGCF.2017.7947610

Pasquero, J. & Bos, J. C. (2017). *System and method for indicating a presence of supplemental information in augmented reality*. US patent: US9685001B2 - System and method for indicating a presence of supplemental information in augmented reality - Google Patents

Pelargos, P. E., Nagasawa, D. T., Lagman, C., Tenn, S., Demos, J. V., Lee, S. J., Bui, T. T., Barnette, N. E., Bhatt, N. S., Ung, N., Bari, A., Martin, N. A., & Yang, I. (2017). Utilizing virtual and augmented reality for educational and clinical enhancements in neurosurgery. *Journal of Clinical Neuroscience*, *35*, 1–4. https://doi.org/10.1016/j.jocn.2016.09.002

Phon, D. N. E., Ali, M. B., & Halim, N. D. A. (2015). Learning with augmented reality: Effects toward student with different spatial abilities. *Advanced Science Letters*, *21*(7), 2200–2204.

Salinas, P. & González-Mendívil, E. (2017). Augmented reality and solids of revolution. *International Journal on Interactive Design and Manufacturing*, *11*(4), 829–837. https://doi.org/10.1007/s12008-017-0390-3

Saltan, F. (2017). *The use of augmentedrReality in formal education: A scopingrReview*. *8223*(2), 503–520. https://doi.org/10.12973/eurasia.2017.00628a

TeKolste, R. D. & Liu, V. K. (2018). *Outcoupling grating for augmented reality system*. U.S. Patent Application 10/073,267. https://patents.google.com/patent/WO2018081305A1/en

Wang, Y. H. (2017). Using augmented reality to support a software editing course for college students. *Journal of Computer Assisted Learning*, *33*(5), 532–546. https://doi.org/10.1111/jcal.12199